\documentclass[aps,pre,showpacs,floatfix,twocolumn,10pt,superscriptaddress]{revtex4-1}

\usepackage[T1]{fontenc}

\usepackage{bm}				
\usepackage{amsmath}
\usepackage{amssymb}
\usepackage{latexsym}
\usepackage{amsfonts}
\usepackage{graphicx}
\usepackage{color}
\usepackage{wasysym}
\usepackage{mathtools}

\newcommand{\be}{\begin{equation}}
\newcommand{\ee}{\end{equation}}
\newcommand{\beq}{\begin{equation}}
\newcommand{\eeq}{\end{equation}}
\newcommand{\la}{\langle}
\newcommand{\ra}{\rangle}
\newcommand{\ben}{\begin{eqnarray}}
\newcommand{\een}{\end{eqnarray}}
\newcommand{\bea}{\begin{eqnarray}}
\newcommand{\eea}{\end{eqnarray}}

\renewcommand{\vec}[1]{{\bf {#1}}}

\newcommand{\vrr}{{\bf{r}}}

\newcommand{\vxi}{{\bm{\xi}}}
\newcommand{\vnabla}{{\bm{\nabla}}}
\newcommand{\vj}{{\bf{j}}}

\newcommand{\vq}{{\bf{q}}}
\newcommand{\vJJ}{\mathbf{J}}

\newcommand{\vlamb}{\bm{\lambda}}

\newcommand{\vecep}{\bm{\epsilon}}

\newcommand{\vla}{\bm{\lambda}}

\newcommand{\zh}{\vlamb}
\newcommand{\zz}{\mathbf{z}}

\newcommand{\wap}{{\text{w}}}
\newcommand{\sap}{{\text{s}}}

\newcommand{\hD}{{\hat{D}}}
\newcommand{\hS}{{\hat{\sigma}}}
\newcommand{\Jpar}{{J_{\parallel}}}
\newcommand{\Jperp}{{\vec{J}_{\perp}}}
\newcommand{\Jperpa}{{J_{\perp}^{(\alpha)}}}
\newcommand{\jpx}{{\vec{j}_{\perp}(x)}}
\newcommand{\jpxo}{{\bar{\vec{j}}_{\wap,\perp}(x;\vJJ)}}
\newcommand{\jpxa}{{j_{\perp}^{(\alpha)}(x)}}
\newcommand{\jpxao}{{\bar{j}_{\wap,\perp}^{(\alpha)}(x;\vJJ)}}

\newcommand{\lperp}{{\bm{\nu}_{\perp}}}
\newcommand{\lpa}{{\nu_{\perp}^{(\alpha)}}}

\def\lam{\lambda}

\def\(({\left(}
\def\)){\right)}
\def\[[{\left[}
\def\]]{\right]}

\begin{document}

\title{Weak additivity principle for current statistics in $d$-dimensions}

\author{C. P\'erez-Espigares}
\email{carlos.perezespigares@unimore.it}
\affiliation{University of Modena and Reggio Emilia, via G. Campi 213/b, 41125 Modena, Italy}
\author{P.L. Garrido}
\email{garrido@onsager.ugr.es}
\author{P.I. Hurtado}
\email{phurtado@onsager.ugr.es}
\affiliation{Institute Carlos I for Theoretical and Computational Physics, and Departamento de Electromagnetismo y F\'isica de la Materia, Universidad de Granada, 18071 Granada, Spain}

\date{\today}


\begin{abstract}
The additivity principle (AP) allows to compute the current distribution in many one-dimensional ($1d$) nonequilibrium systems. Here we extend this conjecture to general $d$-dimensional driven diffusive systems, and validate its predictions against both numerical simulations of rare events and microscopic exact calculations of three paradigmatic models of diffusive transport in $d=2$. Crucially, the existence of a structured current vector field at the fluctuating level, coupled to the local mobility, turns out to be essential to understand current statistics in $d>1$. We prove that, when compared to the straightforward extension of the AP to high-$d$, the so-called weak AP always yields a better minimizer of the macroscopic fluctuation theory action for current statistics.
\end{abstract}

\maketitle


Currents are the hallmark of nonequilibrium behavior. Whenever a system is driven out of equilibrium by a boundary gradient and/or external field, a current of a conjugate observable (mass, energy, momentum, charge, etc.) appears which reflects the associated entropy production \cite{Mazur}. The function controlling current fluctuations seems to play a role akin to the equilibrium free energy in nonequilibrium situations \cite{Touchette,Derrida1}, and hence the understanding of current statistics in terms of microscopic dynamics has become one of the main goals of nonequilibrium statistical mechanics, a problem which has proven very hard even in the simplest situations. Indeed, up to know only a handful exactly-solvable models are fully understood \cite{Derrida1,Derrida2,Lazarescu1,Lazarescu2} and, despite some exact results in the form of fluctuation theorems \cite{ECM,GC,K,LS,Jarzinsky,Crooks,HS,Seifert,weIFR,AFR,AFRexp,wequantum,Gaspard1,Gaspard2,Gaspard3}, the overall picture remains puzzling and in need of a general, first-principles approach. This deadlock has changed dramatically with the recent formulation of macroscopic fluctuation theory (MFT) \cite{Bertini1,Bertini2,Bertini3,Bertini4,Bertini5,Bertini6,Bertini7,Bertini8,Bertini9}, an unifying theoretical scheme to study dynamic fluctuations in nonequilibrium systems, based solely on the knowledge of a few transport coefficients easily measurable in experiments, and applicable to a broad class of nonequilibrium problems \cite{weSSB1,weSSB2,Jack,wePLH1,wePLH2,wePLH3,BL,singlefile,absorption,melting,Bouchet,Tailleur,Derrida3,Gerschenfeld,Krapivsky,Meerson1,Meerson2,Meerson3,wewave}.

When applied to current statistics, MFT leads to a well-defined but highly-complex variational problem in space and time for the optimal paths responsible of a given current fluctuation, whose solution remains challenging in most cases \cite{Bertini3,Bertini4,Bertini5,Bertini6,Bertini7}. However, in an effort to explore clarifying hypotheses, Bodineau and Derrida \cite{BD} (see also \cite{Derrida1,Derrida2,Bertini7}) have conjectured an additivity principle (AP) which greatly simplifies the MFT variational problem for currents in $1d$, leading to explicit quantitative predictions and thus opening the door to a systematic way of computing the current statistics in general nonequilibrium systems \cite{Touchette}. In few words, the AP amounts to assuming within MFT that the optimal path responsible of a given current fluctuation is time-independent. The validity of the AP has been confirmed with high accuracy in rare-event simulations of $1d$ stochastic lattice gases \cite{weAddit,weAdditPRE,Gorissen,weJSP}, but the question remains however as to how to generalize this conjecture to the more interesting case of $d>1$.

Here we propose such a generalization, that we call weak additivity principle (wAP), and demonstrate its validity and accuracy by comparing our predictions with both numerical simulations of rare events \cite{sim1,sim2,sim3,sim4,weJSTAT} and microscopic exact calculations \cite{AFR,Schutz1,Schutz2,Schutz3,CarlosSFR} in three paradigmatic models of diffusive transport, namely the Kipnis-Marchioro-Presutti (KMP) model of heat conduction \cite{kmp}, the Zero-Range Process (ZRP) \cite{zrp,MukSch} and the Random Walk (RW) model \cite{Spohn,Bertini7}, all defined in $d=2$. A main novelty of our conjecture when compared to the straightforward generalization of the $1d$ AP to $d>1$ is the realization of the essential role played by an optimal divergence-free current vector field in the MFT variational problem for current statistics in $d>1$. This optimal current field turns out to be structured along the gradient direction according to the local mobility, a possibility already suggested in \cite{AFR}. It is then easy to prove that the wAP always yields a better minimizer of the MFT action for current statistics. 

We are interested in a broad class of $d$-dimensional driven diffusive systems characterized by a conserved density field $\rho(\vrr,t)$ which evolves according to the following fluctuating hydrodynamics equation \cite{Bertini3,Bertini4,Bertini5,Bertini6,Bertini7,Derrida1,Derrida2,weJSP}
\be
\partial_t \rho(\vrr,t) + \vnabla \cdot \left( -\hD(\rho) \vnabla \rho(\vrr,t) + \vxi(\vrr,t) \right) = 0 \, ,
\label{langevin}
\ee
with $\vrr\in \Lambda\equiv [0,1]^d$. The field $\vj (\vrr,t)\equiv -\hD(\rho) \vnabla \rho(\vrr,t) + \vxi(\vrr,t)$ is the fluctuating current, with local average given by Fick's/Fourier's law with a diffusivity matrix $\hD(\rho)$, and $\vxi(\vrr,t)$ is a Gaussian white noise with $\la \vxi(\vrr,t)\ra=0$, and characterized by a mobility matrix $\hS(\rho)$ 
\be
\la \xi_\alpha(\vrr,t)\xi_\beta(\vrr',t') \ra=L^{-d}\sigma_\alpha(\rho)\delta_{\alpha\beta}\delta(\vrr-\vrr')\delta(t-t') \, , \nonumber
\ee
with $L$ the system size in natural units and $\alpha,\beta\in [1,d]$. This (conserved) noise term accounts for the many fast microscopic degrees of freedom which are averaged out in the coarse-graining procedure resulting in Eq. (\ref{langevin}). The diffusion and mobility transport matrices are diagonal, with components $D_\alpha(\rho)$ and $\sigma_\alpha(\rho)$ respectively, being related via a local Einstein relation $\hD(\rho)=f_0''(\rho) \hS(\rho)$, with $f_0(\rho)$ the \emph{equilibrium} free energy of the system at hand. To completely define the problem, the evolution equation (\ref{langevin}) must be supplemented with appropriate boundary conditions, which typically include an external gradient along a given direction (say $\hat{x}$), $\rho(\vrr,t)\vert_{x=0,1}=\rho_{L,R}$, which drives the system out of equilibrium for $\rho_L\neq \rho_R$, and periodic boundaries along all other $(d-1)$ directions.  

The probability of observing a given history $\{\rho(\vrr,t),\vj(\vrr,t)\}_0^{\tau}$ of duration $\tau$ for the density and current fields can be written using path integrals as \cite{Bertini7}
\be
\text{P}\left(\{\rho,\vj\}_0^{\tau} \right) \sim \exp \Big( +L^d I_{\tau}\left[\rho,\vj \right] \Big) \, , \nonumber 
\ee
with an action $I_{\tau}\left[\rho,\vj \right]=- \int_0^{\tau} dt \int_\Lambda d \vrr\, {\cal L} (\rho,\vj)$ and
\be
{\cal L} (\rho,\vj)= \frac{1}{2}\left(\vj+\hD(\rho) \vnabla\rho \right) \cdot \hat{\Sigma}(\rho) \left(\vj+\hD(\rho) \vnabla\rho \right) \,  . \nonumber
\ee
The matrix $\hat{\Sigma}(\rho)$ is diagonal with components $\Sigma_\alpha(\rho)\equiv \sigma_\alpha^{-1}(\rho)$, and the fields $\rho (\vrr,t)$ and $\vj(\vrr,t)$ are coupled via the continuity equation, see also Eq. (\ref{langevin})
\be
\partial_t \rho (\vrr,t) + \vnabla\cdot \vj(\vrr,t) = 0 \, . 
\label{conteq}
\ee
In any other case $I_{\tau}\left[\rho,\vj \right]\to-\infty$. The probability $\text{P}_{\tau}(\vJJ)$  of observing an averaged empirical current $\vJJ$, defined as
\be
\vJJ = \frac{1}{\tau}  \int_0^{\tau} dt \int_\Lambda d\vrr \, \, \vj (\vrr,t) \, ,
\label{empJ}
\ee
scales for long times as $\text{P}_{\tau}(\vJJ)\sim \exp[+\tau L^d G(\vJJ)]$, and the current large deviation function (LDF) $G(\vJJ)$ can be related to $I_{\tau}[\rho,\vj]$ via a simple saddle-point calculation in the long-time limit, $G(\vJJ) = \lim_{\tau \to \infty} \tau^{-1} \max_{\{\rho,\vj\}} I_{\tau}[\rho,\vj]$, subject to constraints (\ref{conteq}) and (\ref{empJ}) and the fixed boundary conditions. The density and current fields solution of this variational problem, denoted here as $\bar\rho(\vrr,t;\vJJ)$ and $\bar\vj(\vrr,t;\vJJ)$, are just the optimal path the system follows to sustain a long-time current fluctuation $\vJJ$. 

This is a complex spatiotemporal variational problem whose solution remains challenging in most cases \cite{Derrida1,Derrida2,Bertini3,Bertini4,Bertini5,Bertini6,Bertini7,BD,weAddit,weAdditPRE,weJSP,BD2,weSSB1,weSSB2}, so simplifying hypotheses are required. Inspired by results from $1d$ \cite{Derrida1,Derrida2,BD,weAddit,weAdditPRE,weJSP}, we now propose a weak version of the additivity principle (or wAP in short) which consists in two main hypotheses, namely that (i) the dominant paths responsible for a given current fluctuation are indeed time-independent \cite{note1}, i.e. $\rho(\vrr)$ and $\vj(\vrr)$, and (ii) the relevant fields exhibit structure only along the gradient direction, so $\rho(x)$ and $\vj(x)$ in our convention. Clearly (ii) is expected on physical grounds due to periodicity along all directions orthogonal to the gradient. To make clear the simplifying power of the wAP, note that (i) implies, via the continuity equation (\ref{conteq}), that the relevant current vector fields are divergence-free, $\vnabla\cdot \vj(\vrr)=0$, and this, together with (ii) above and constraint (\ref{empJ}), leads to current fields $\vj(x)=\left(\Jpar,\jpx\right)$, with
\be
\Jperp=\int_0^1 dx \, \jpx \, ,
\label{Jperp}
\ee
and where we have decomposed $\vJJ=(\Jpar,\Jperp)$ along the gradient ($\parallel$) and all other, $(d-1)$ directions ($\perp$). The wAP thus leads to the following simplified variational problem for the current LDF
\ben
G_\wap(\vJJ)&=&-\min_{\rho(x) \atop \jpx} \int_0^1 dx \, {\cal L}_\wap(\rho,\vec{j}_\perp;\vJJ) \, ,  \label{ldfwAP} \nonumber \\
{\cal L}_\wap(\rho,\vec{j}_\perp;\vJJ)&=&\frac{[\Jpar+D_1(\rho)\rho'(x)]^2}{2\sigma_1(\rho)} + \sum_{\alpha=2}^d \frac{\jpxa^2}{2\sigma_\alpha(\rho)} \, , \label{lagrangianwAP} \nonumber
\een
and subject to the constraints (\ref{Jperp}) and the imposed boundary conditions. To explicitly take into account the constraints, we now introduce $(d-1)$ Lagrange multipliers and define a modified functional ${\cal L}_\wap^{(\lperp)}(\rho,\vec{j}_\perp;\vJJ)\equiv {\cal L}_\wap(\rho,\vec{j}_\perp;\vJJ) - \lperp\cdot\jpx$. Standard variational calculus thus leads to the following differential equation for the optimal density profile $\bar\rho_\wap(x;\vJJ)$ \cite{weJSP}
\be
D_1(\rho)^2 \rho'(x)^2=\Jpar^2 + \sigma_1(\rho) \left[ 2K - \sum_{\alpha=2}^d \lpa^2 \sigma_\alpha(\rho) \right] \, , \nonumber
\ee
where $K$ is an integration constant which guarantees the correct boundary conditions \cite{weJSP}. The optimal current  field also follows as $\bar{\vj}_\wap(x;\vJJ)=\left(\Jpar,\jpxo\right)$ with
\be
\jpxao=\lpa\sigma_\alpha(\bar\rho_\wap) \, , \quad \alpha\in [2,d] \, ,
\label{jopt}
\ee
with the Lagrange multipliers fixed via (\ref{Jperp}) to $\lpa =  \Jperpa / \int_0^1 dx \, \sigma_\alpha(\bar\rho_\wap)$. Eq.~(\ref{jopt}) shows that the optimal, divergence-free current vector field exhibits structure along the gradient direction in all orthogonal components, and this structure is coupled to the optimal density profile via the mobility transport coefficient. 

This result should be compared with the straightforward extension of the $1d$-AP to high dimensions, which amounts to assume, together with (i)-(ii) above, that the optimal current field is constant across space and hence equals $\vJJ$ due to (\ref{empJ}). This strong additivity principle (or sAP in short) leads to an even simpler variational problem for the current LDF, $G_\sap(\vJJ)=-\min_{\rho(x)} \int_0^1 dx \, {\cal L}_\sap(\rho;\vJJ)$, with
\be
{\cal L}_\sap(\rho;\vJJ)=\frac{[\Jpar+D_1(\rho)\rho'(x)]^2}{2\sigma_1(\rho)} + \sum_{\alpha=2}^d \frac{{J_\perp^{(\alpha)}}^2}{2\sigma_\alpha(\rho)} \, , \label{lagrangiansAP} \nonumber
\ee
whose optimal solution is denoted here as $\bar\rho_\sap(x;\vJJ)$. Note that, for $\vJJ$ fixed, we expect $\bar\rho_\sap(x;\vJJ) \ne \bar\rho_\wap(x;\vJJ)$ in general, and the question remains as to which hypothesis (wAP or sAP) yields a maximal $G(\vJJ)$. Intuition suggests that the wAP should offer a better solution as it includes additional degrees of freedom that the system at hand can \emph{put at work} to improve its rate function. To confirm rigorously this argument, note first that the optimal current field $\bar{\vj}_\wap(x;\vJJ)$ is a functional of the optimal density $\bar\rho_\wap(x;\vJJ)$, see Eq. (\ref{jopt}), so we can always write $G_\wap(\vJJ)={\cal F}_{\wap}(\bar\rho_\wap;\vJJ)$, where we have defined the functional ${\cal F}_\ell(\psi;\vJJ)\equiv -\int_0^1 dx {\cal L}_\ell(\psi;\vJJ)$, with $\ell=\text{w, s}$, for any function $\psi(x)$ obeying the boundary conditions. Similarly, we may write $G_\sap(\vJJ)={\cal F}_\sap(\bar\rho_\sap;\vJJ)$. Since $\bar\rho_\wap(x;\vJJ)$ is the maximizer of the wAP action, clearly  ${\cal F}_{\wap}(\bar\rho_\wap;\vJJ) \ge {\cal F}_\wap(\psi;\vJJ)$ $\forall \psi(x)\ne \bar\rho_\wap(x;\vJJ)$. Next, we compare both functionals ${\cal F}_{\wap,\sap}$ applied to \emph{the same} profile $\bar\rho_\sap$ at fixed $\vJJ$, i.e. we define $\Delta_{\wap\sap}\equiv {\cal F}_{\wap}(\bar\rho_\sap;\vJJ) - {\cal F}_{\sap}(\bar\rho_\sap;\vJJ)$ and find
\be
\Delta_{\wap\sap}= \sum_{\alpha=2}^d \frac{\Jperpa^2}{2} \left[ \int_0^1 dx \, \frac{1}{\sigma_\alpha(\bar\rho_\sap)}  - \frac{1}{ \int_0^1 dx \, \sigma_\alpha(\bar\rho_\sap)}\right]  \ge 0 . \nonumber
\ee
The last inequality arises because $\int_0^1 dx \sigma_\alpha^{-1}(\bar\rho_\sap) \ge (\int_0^1 dx \sigma_\alpha(\bar\rho_\sap))^{-1}$, which is a particular instance of the reverse H\"older's inequality \cite{Holder}. In this way, ${\cal F}_{\wap}(\bar\rho_\wap;\vJJ) \ge {\cal F}_{\wap}(\bar\rho_\sap;\vJJ) \ge {\cal F}_{\sap}(\bar\rho_\sap;\vJJ)$  and hence $G_\wap(\vJJ) \ge G_\sap(\vJJ)$. This proves that, when compared to the strong AP, the weak AP always yields a better minimizer of the macroscopic fluctuation theory action for currents. This result therefore singles out the wAP as the relevant simplifying hypothesis to study current statistics in general $d$-dimensional systems. Interestingly, the previous proof shows that both the sAP and wAP yield the same result only for constant mobility, $\sigma_\alpha(\rho)=\sigma_\alpha$ $\forall \alpha$, or for current fluctuations parallel to the gradient direction, $\vJJ=(\Jpar,\Jperp=0)$. This observation helps in making sense of previous, seemingly contradictory results \cite{APdDhar,APdDerrida,APdCleuren}.

\begin{figure}
\vspace{-0.3cm}
\includegraphics[width=8.5cm]{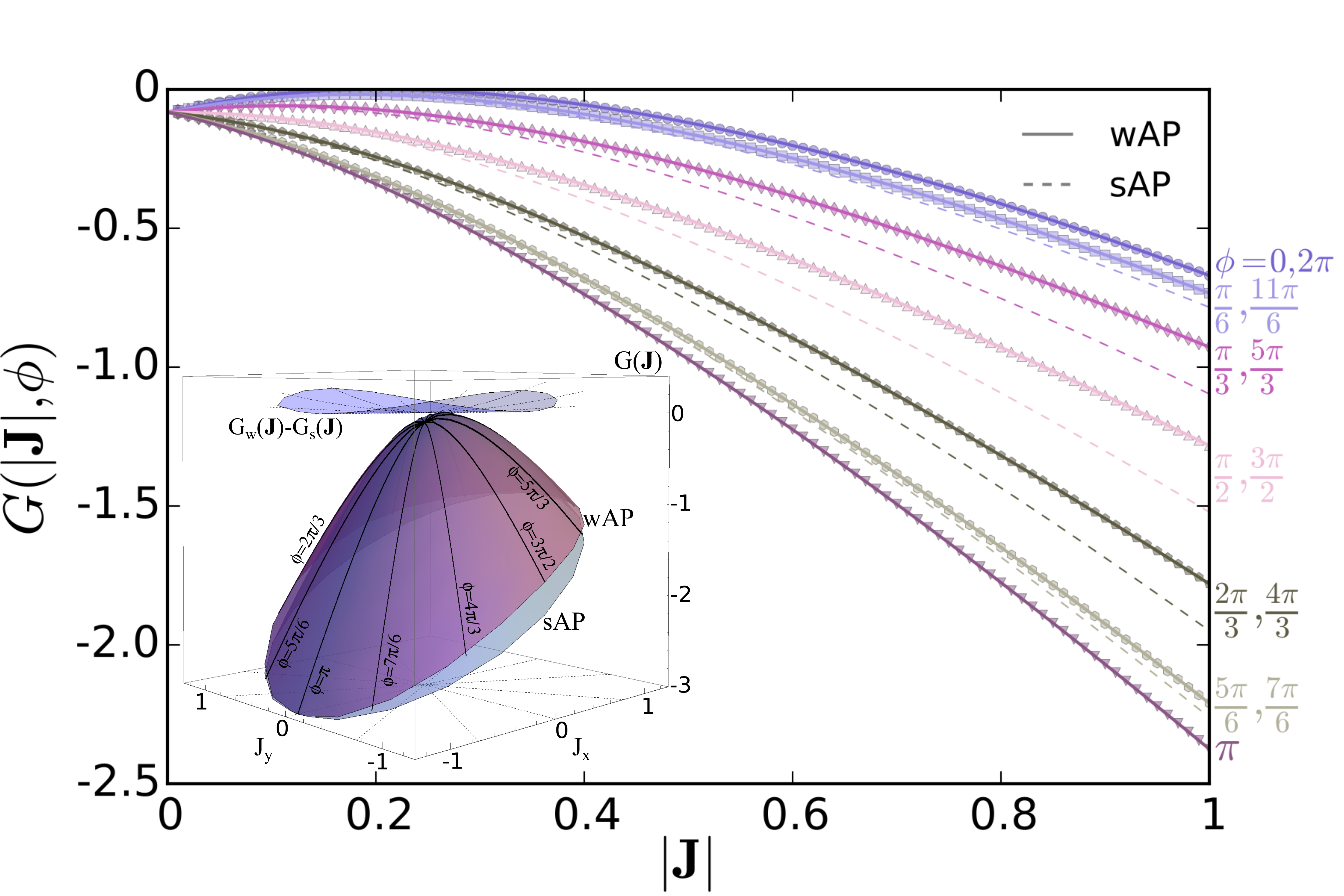}
\includegraphics[width=8.5cm]{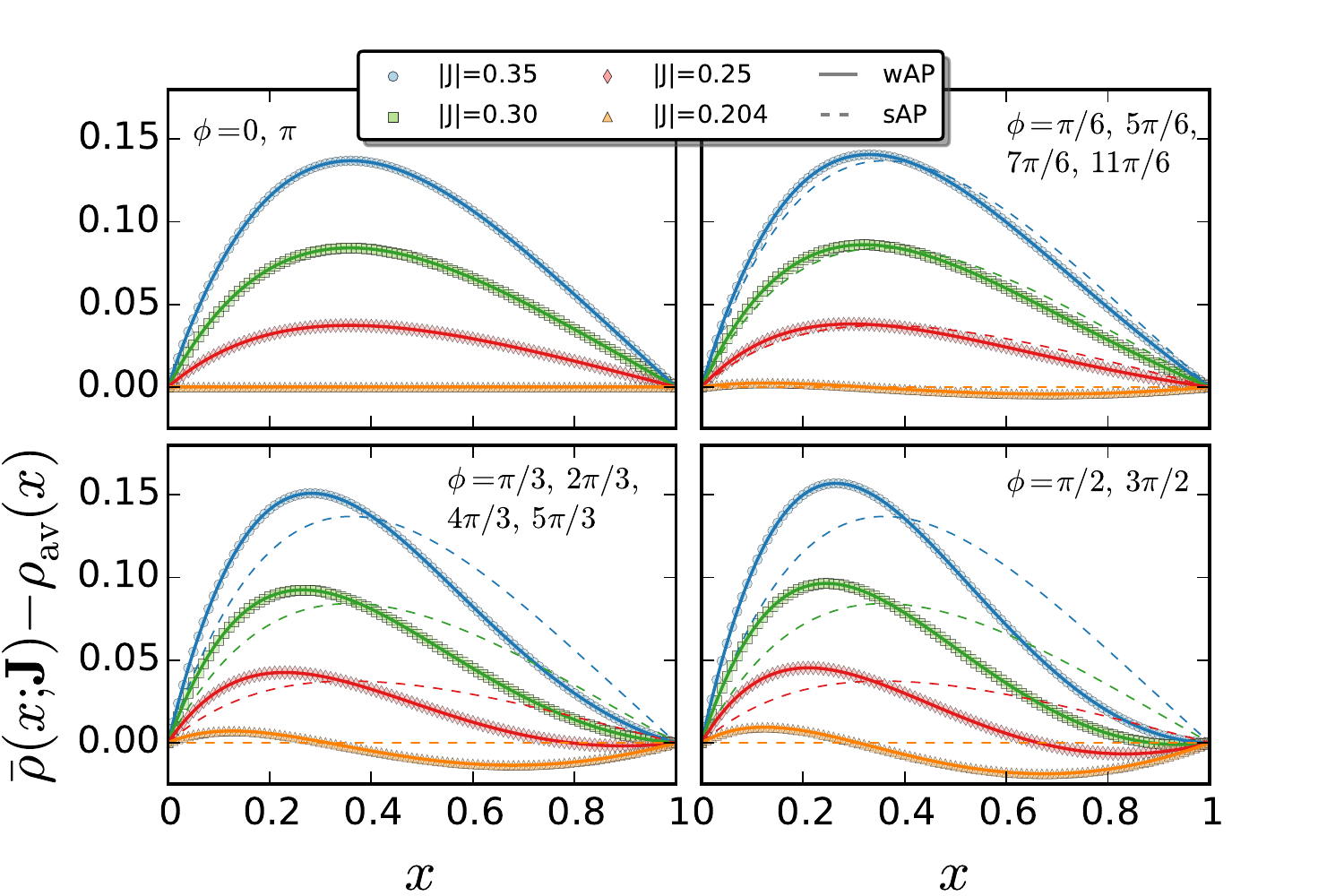}
\vspace{-0.2cm}
\caption{\small (Color online) Top: Current LDF for the isotropic ZRP vs $|\vJJ|$ for different angles $\phi=\tan^{-1}(J_y/J_x)$. Inset: $G(\vJJ)$  from MFT under wAP and sAP. Clearly, $G_\wap(\vJJ)\ge G_\sap(\vJJ)$. Bottom: Excess optimal density profiles for different $|\vJJ|$ and $\phi$. Symbols stand for exact matrix computations for $L=10^5$, while solid (dashed) lines represent wAP (sAP) predictions. 
}
\label{fig1}
\end{figure}

Our aim now is to verify the wAP predictions against both numerical simulations of rare events and microscopic exact calculations of various paradigmatic models of diffusive transport in $d=2$. Our first model of choice is the widely-studied Zero Range Process (ZRP) \cite{zrp,MukSch}, a model of interacting particles amenable to exact computations due to the factorization property of its stationary measure. The ZRP is defined on a $d$-dimensional lattice of linear size $L$ whose sites $i$ may be occupied by an arbitrary number of particles $n_i\in\mathbb{N}$ which jump to randomly chosen neighbors at a rate $\omega_\alpha(n_i)=h_\alpha f(n_i)$, with $f(n_i)$ the interaction function (which depends only on the population of the departure site) and $h_\alpha$ the (constant) hopping rate along the $\alpha$-direction, $\alpha\in [1,d]$. Different interaction functions model varying physical situations, but for concreteness we focus here on a constant $f(n)=1$, which mimicks an effective attraction between particles on each site \cite{zrp}. When coupled to particle reservoirs at the left and right boundaries at densities $\rho_L$ and $\rho_R$ respectively \cite{zrp,MukSch}, with $\rho_L\ne \rho_R$, the so-defined ZRP sustains a net average current of particles $\la \vJJ\ra = \hat{x} h_1 (\rho_L-\rho_R)/[(1+\rho_L)(1+\rho_R)]$ described by Fick's law with a diffusivity matrix with components $D_\alpha(\rho)=h_\alpha /(1+\rho)^{2}$. Moreover, the mobility coefficient has components $\sigma_\alpha(\rho)=2h_\alpha \rho/(1+\rho)$, and together these transport coefficients can be used to solve numerically the MFT problem for currents under the wAP conjecture (see \cite{Schutz3} for the 1$d$ case). We compare these theoretical predictions with exact results for the ZRP current LDF and the associated optimal density profiles, that can be obtained within the so-called quantum Hamiltonian formalism for the master equation \cite{AFR,Schutz1,Schutz2,Schutz3}. Within this picture, the current LDF is obtained from the lowest eigenvalue of a \emph{tilted Hamiltonian}, a spectral problem which reduces to a $L\times L$ system of linear equations due to the factorization property of ZRP \cite{AFR,Schutz1,Schutz2,Schutz3}, see Appendix A. Optimal density profiles are then related to the left and right eigenvectors associated to the lowest eigenvalue \cite{weJSTAT,weJSP,weAdditPRE}. Fig.~\ref{fig1} shows our results for $G(\vJJ)$ (top) and $\bar\rho(x;\vJJ)$ (bottom, after subtracting the steady-state profile $\rho_\text{av}(x)$ \cite{rhoav}) for parameters $\rho_L=1$, $\rho_R=0.1$, and isotropic hopping rates $h_\alpha=1/2\, \forall \alpha$. The agreement between wAP predictions and exact matrix computations for $L=10^5$ is excellent in all cases, while sAP predictions fail outside the gradient direction, the discrepancy being maximal for orthogonal fluctuations and increasing with $|\vJJ|$. Appendix B presents similar data for an anisotropic ZRP, as well as for a fluid of random walkers, and in all cases the agreement between wAP predictions and matrix data for $L=10^5$ is remarkable.

\begin{figure}
\vspace{-0.3cm}
\includegraphics[width=8.5cm]{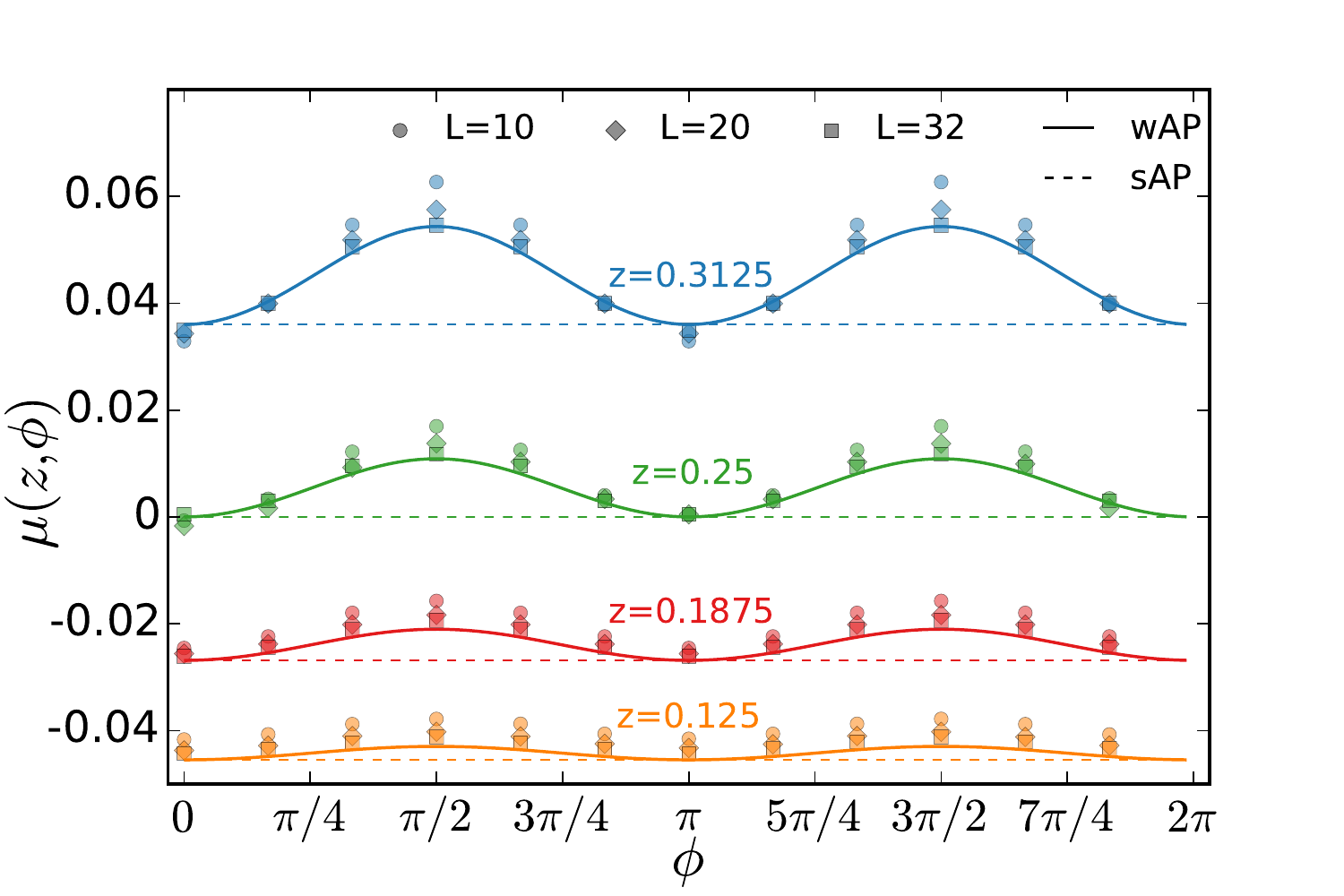}
\hspace{0.2cm}\includegraphics[width=8.8cm]{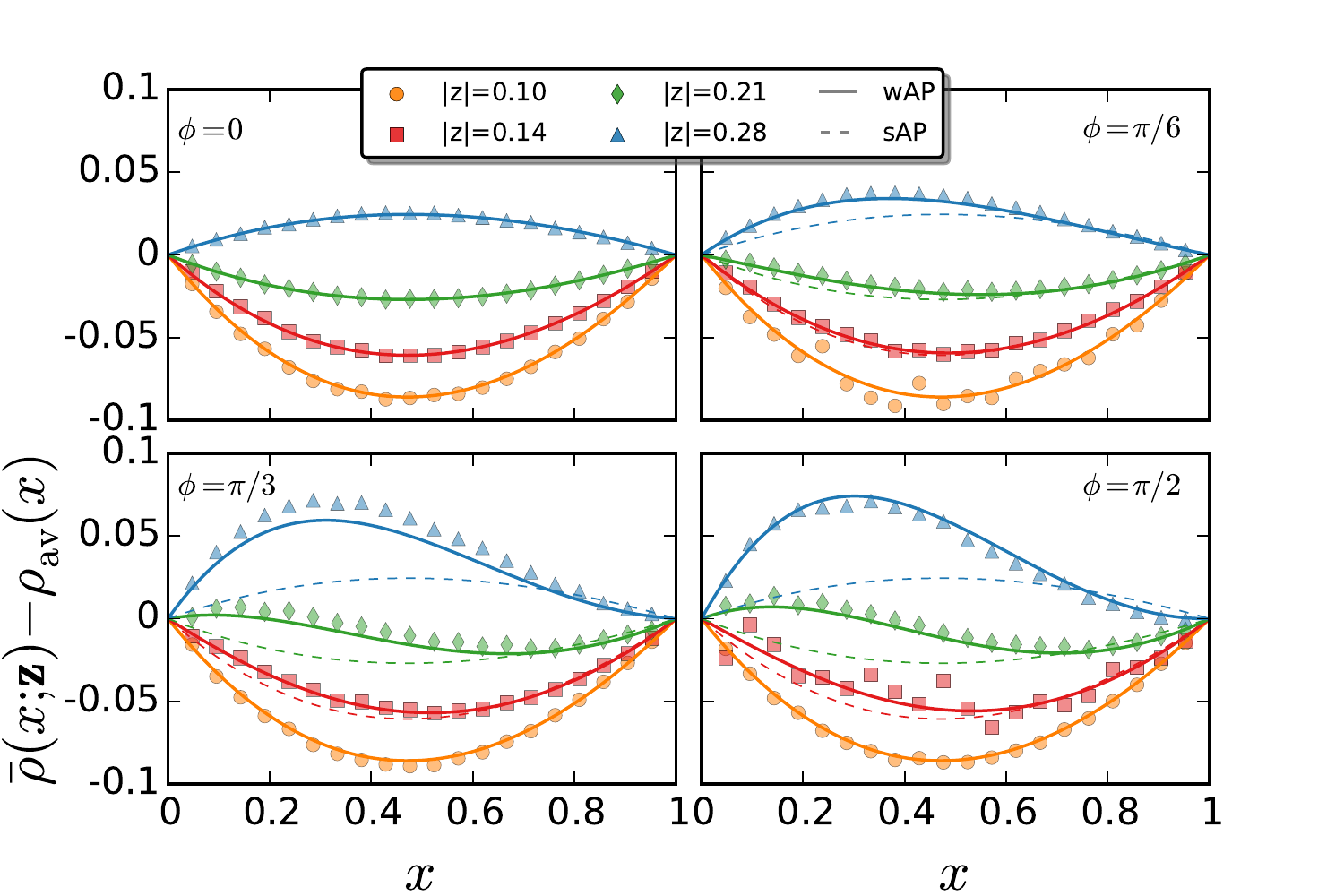}
\vspace{-0.5cm}
\caption{\small (Color online) Top: Legendre transform of the current LDF for the KMP model vs $\phi$ for different values of $z\equiv |\zz|$ and varying $L$.  Convergence to the wAP prediction as $L$ increases is apparent. Bottom: Excess optimal density profiles for different $z$ and $\phi$ as measured for $L=20$. Symbols stand for cloning simulation results, while solid (dashed) lines represent wAP (sAP) predictions. 
}
\label{fig2}
\end{figure}

The previous results are restricted to transport models with a factorizable stationary measure \cite{zrp}. We now focus on the more complex $2d$-KMP model of heat transport \cite{kmp}, defined on a square lattice of linear size $L$ whose sites $i$ contain certain amount of energy $\rho_i\in \mathbb{R}_+$. Dynamics proceeds via random energy exchanges between neighbors, such that the pair energy is conserved, and we couple the system to two thermal baths at the left and right ends at temperatures $T_{L,R}$, respectively \cite{kmp,weJSP}, with periodic boundary conditions in the $y$-direction. At the macroscopic level this model obeys Fourier's law with a scalar conductivity $D(\rho)=1/2$ and a mobility $\sigma(\rho)=\rho^2$, and for $T_L\ne T_R$ it develops a linear temperature profile $\rho_\text{av}(x)=T_L + x\,(T_R-T_L)$ with a nonzero average current $\la \vJJ\ra = \hat{x}(T_L-T_R)/2$. For this non-factorizable model the quantum Hamiltonian matrix approach does not yield useful results. Instead, we measure the full current statistics using advanced cloning Monte Carlo simulations particularly designed for this task \cite{sim1,sim2,sim3,sim4,weIFR,weJSP,weAdditPRE}. This method, which works well for not too large $L$, yields the Legendre-Fenchel transform of the current LDF, $\mu(\zh)=\max_\vJJ[G(\vJJ) + \zh\cdot \vJJ]$. Fig.~\ref{fig2} shows the measured $\mu(\zh)$ for $T_L=2$, $T_R=1$ and different $L$, as a function of the current angle $\phi$ for different values of $|\zz|$, with $\zz\equiv \zh+\vecep$ and $\vecep=\frac{1}{2}(T_R^{-1}-T_L^{-1})$, corresponding to a broad range of current fluctuations \cite{weIFR}. While the sAP predicts a $\phi$-independent $\mu(\zh)$ for fixed $|\zz|$, we observe a double-bump structure in $\phi$ as predicted by wAP \cite{weNico}. Moreover, finite-size data clearly converge to the wAP prediction as $L$ increases, while sAP only yields the correct prediction for $\phi=0,\pi$, as expected. Note that similar finite-size corrections are observed for the ZRP, see Appendix B. Data for optimal density profiles also fit nicely the theoretical wAP curves, overall demonstrating the superior predictive power of the weak additivity principle presented in this paper. 

In summary, we have extended the additivity principle to general $d$-dimensional driven diffusive systems, demonstrating the key role played by a structured current field (coupled to the local density via the mobility coefficient) to understand current statistics in $d>1$. Predictions from the so-called weak additivity principle have been tested against both exact matrix results and simulations of rare events in different paradigmatic models of transport in $d=2$, and a remarkable agreement is found in all cases. Moreover, we have also proven that the wAP (and not the sAP) offers a better minimizer of the MFT action for currents, except for current fluctuations along the gradient direction, where both wAP and sAP yield equivalent results. This explains previous apparent validations of the sAP in $d$-dimensional systems \cite{APdDhar,APdDerrida,APdCleuren}, as these works focus on a scalar current parallel to the gradient. However, in the general vectorial-current case the role of the structured, divergence-free optimal current field associated to the wAP cannot be overlooked. Indeed, our general findings agree with very recent microscopic results for the ZRP which highlight the importance of the local structure of the current field in this model \cite{Rodrigo}. An interesting issue for future study concerns the stability of the wAP solution against space\&time perturbations in $d$-dimensional boundary driven systems \cite{Akkermans}. Finally, we mention that additivity violations are known to happen in $1d$ periodic systems via a dynamic phase transition to a traveling-wave phase with broken symmetries \cite{BD2,weSSB1,weSSB2,Jack,Bertini4,Bertini7}. The natural question of course concerns the nature of this transition for $d>1$. We anticipate that a similar spontaneous symmetry-breaking phenomenon exists at the fluctuation level in $d$-dimensions, for which a form of weak additivity in terms of a structured current field also plays a crucial role \cite{weNico}.

We thank R.J. Harris, N. Tiz\'on and R. Villavicencio-S\'anchez for useful discussions. Financial support from Spanish project FIS2013-43201-P (MINECO), NSF grant DMR1104500, Italian Research Funding Agency (MIUR) through FIRB project grant RBFR10N90W, Italian INdAM \emph{Francesco Severi}, University of Granada, Junta de Andaluc\'{\i}a project P09-FQM4682 and GENIL PYR-2014-13 project is acknowledged.

\appendix

\onecolumngrid

\section{Quantum Hamiltonian formalism for the current statistics of the $2d$ ZRP}
\label{apA}

In this appendix we derive in a self-contained way the current statistics of the $2d$ Zero-Range Process (ZRP) using the quantum Hamiltonian formalism for the master equation as main tool. In particular, we follow Ref. \cite{Rodrigo} where a similar calculation has been recently presented. Our aim is to calculate the current scaled cumulant generating function $\mu_L(\vla)$ and the microscopic optimal density profiles associated to a given current fluctuation in a system of linear size $L$. Within the quantum Hamiltonian formalism, the master equation is written in Schr\"odinger form \cite{QHF1,QHF2} as  
\be
\frac{d|P\ra}{dt}=-H|P\ra \, ,
\ee
with the so-called \emph{Hamiltonian} $H$ given by the stochastic generator containing the transition rates between all states of the system. We have introduced Dirac's bra and ket notation, with the ket $|P\ra$ representing the probability column vector $(P(C_1),P(C_2),\dots)^T$, with $^T$ denoting transposition, and where $P(C_k)$ denotes the probability measure on the set of all configurations $C_k=(n_1,n_2,...,n_M)$, $n_i\in \mathbb{N}$, being $n_i$ the number of particles on site $i$ (out of a total number of $M$ sites). The probability vector is then defined as $|P\ra=\sum_k P(C_k) |C_k\ra$ where $|C_k\ra$ is a basis vector for the particle configuration, i.e. it corresponds to the column vector $|C_k\ra=(0,0,\dots,0,1,0,\dots)^T$ with all components equal to zero except for the component corresponding to configuration $C_k$. The probability vector is normalized such that $\la 1|P\ra=1$ where $\la 1|=\sum_k \la C_k|$ is the row vector with all elements equal to one and $\la C|C'\ra=\delta_{CC'}$. One can readily verify that $\la 1|$ is the left-eigenvector of $H$ with zero eigenvalue $\la 1|H=0$ (expressing conservation of probability). On the other hand, the stationary distribution or {\em ground state} of the stochastic process, denoted here as $|P^*\ra$, corresponds to the right-eigenvector of $H$ with zero eigenvalue
\be
H|P^*\ra=0.
\ee

The $2d$ ZRP we consider here is defined on a square lattice of linear size $L$ with particle reservoirs at the boundaries in the $x$-direction and with periodic boundary conditions in the $y$-direction. Configurations are denoted as $C_k=(n_{11},n_{12},...,n_{LL})$, $n_{ji}\in \mathbb{N}$, $j,i\in[1,L]$, being $n_{ji}$ the number of particles on site $(j,i)$. Notice that for each site in the square lattice, $j$ denotes the row index while $i$ denotes the column. The dynamics is as follows: In the bulk, particles jump to randomly chosen nearest neighbors at a rate $\omega_\alpha(n_{ji})=h_\alpha f(n_{ji})$, with $f(n_{ji})$ the interaction function (which depends only on the population of the departure site) and $h_\alpha$ the (constant) hopping rate along the $\alpha$-direction ($x$ or $y$-direction). In addition, particles are injected at rate $\alpha$ (and removed at rate $\gamma$) at the left boundary -corresponding to the first column of sites- and injected at rate $\delta$ (and removed at rate $\beta$) at the right boundary -corresponding to the last column. Notice that anisotropy can be modeled in this model by considering $h_x\neq h_y$.

As for the one dimensional ZRP with open boundaries \cite{MukSch}, the stationary distribution is given by a product measure
\be
\label{statmea}
|P^*\ra=|P^*_{1,1}) \otimes |P^*_{1,2}) \otimes \cdots \otimes |P^*_{L,L})
\ee
where $|P^*_{j,i})$ is the probability vector corresponding to the marginal distribution for the site $(j,i)$, i.e, $|P^*_{j,i})= \sum_{n_{ji}}P^*_{j,i}(n_{ji})|n_{ji})$, whose components correspond to the probability of finding $n_{ji}$ particles on site $(j,i)$:
\be
\label{probnk}
P_{j,i}^*(n_{ji})=\frac{z_{j,i}^{n_{ji}}}{Z_{j,i}}\prod_{k=1}^{n_{ji}} f(k)^{-1}.
\ee
Here $z_{j,i}$ is the fugacity of site $¡(j,i)$ and $Z_{j,i}$ is the local analogue of the grand-canonical partition function
\be
\label{partfunc}
Z_{j,i}\equiv Z(z_{j,i})=\sum_{n=0}^\infty z_{j,i}^n\prod_{k=1}^{n}f(k)^{-1}.
\ee
It is important to note that the convergence of the partition function depends on how we choose the interaction function $f(k)$. In this work, we restrict to the ZRP in the fluid regime (i.e. in the absence of condensation), so for the $f(k)$'s chosen (either constant or proportional to the number of particles) we consider below current fluctuations within the radius of convergence of \eqref{partfunc}.  In order to relate the fugacity and the mean density on site $(j,i)$, we introduce now the local particle number operator as the diagonal matrix ${\hat n}_{ji}$ with diagonal elements $n_{ji}$. Notice that ${\hat n}_{ji}$ acts exclusively on the $(j,i)^{th}$ component of the configuration vector. Then by using \eqref{probnk} and \eqref{partfunc}, we find 
\be
\rho_{j,i}\equiv \la n_{ji} \ra=\la 1|{\hat n}_{ji}|P^*\ra=\sum_{n_{ji}=0}^{\infty}n_{ji} P^*_{j,i}(n_{ji})=z_{j,i}\frac{\partial \log Z_{j,i}}{\partial z_{j,i}}.
\ee

In order to write explicitly the Hamiltonian of the $2d$ ZRP, we introduce now the following ladder and diagonal operators,
\bea
a_{ji}^+=\left( {\begin{array}{ccccc}
   0 & 0 & 0 & 0 & \cdots \\
   1 & 0 & 0 & 0 & \cdots \\
   0 & 1 & 0 & 0 & \cdots \\
   0 & 0 & 1 & 0 & \cdots \\
   \cdots & \cdots & \cdots & \cdots & \cdots \\
  \end{array} } \right),~~~
a_{ji}^-=\left( {\begin{array}{ccccc}
   0 & f(1) & 0 & 0 & \cdots \\
   0 & 0 & f(2) & 0 & \cdots \\
   0 & 0 & 0 & f(3) & \cdots \\
   0 & 0 & 0 & 0 & \cdots \\
   \cdots & \cdots & \cdots & \cdots & \cdots \\
  \end{array} } \right)~~~
d_{ji}=\left( {\begin{array}{ccccc}
   0 & 0 & 0 & 0 & \cdots \\
   0 & f(1) & 0 & 0 & \cdots \\
   0 & 0 & f(2) & 0 & \cdots \\
   0 & 0 & 0 & f(3) & \cdots \\
   \cdots & \cdots & \cdots & \cdots & \cdots \\
  \end{array} } \right).
  \eea
The subscript ($j,i$) indicates that the respective matrix acts non-trivially only on site ($j,i$) of the lattice, and as a unit operator on all other sites. In this way, the Hamiltonian of the $2d$ ZRP in a square lattice reads \cite{Rodrigo} 
\bea
\label{ham}
-H&=&\sum_{j=1}^L\Big\{\alpha (a_{j,1}^+-1)+\gamma (a_{j,1}^--d_{j,1}) + \delta (a_{j,L}^+-1)+\beta (a_{j,L}^--d_{j,L})\nonumber\\
&+&\sum_{i=1}^{L-1} h_x (a_{j,i}^-a_{j,i+1}^+ -d_{j,i}) + h_x (a_{j,i}^+a_{j,i+1}^- -d_{j,i+1})\nonumber\\
&+&\sum_{i=1}^L h_y (a_{j,i}^-a_{j+1,i}^+ -d_{j,i}) + h_y (a_{j,i}^+a_{j+1,i}^- -d_{j+1,i})\Big\},
\eea
Note that, due to the periodic boundary conditions along the $y$-direction, we identify $j=L+1$ with $j=1$. The first line of the r.h.s of the above equation reflects the injection and extraction of particles from the boundary reservoirs, i.e. it corresponds to the boundary pairs in the first and last column. The second and the third lines correspond to the interaction of the $L(L-1)$ horizontal and the $L^2$ vertical bulk pairs, respectively.

\subsection{Current fluctuations for the $2d$ ZRP}

Our first task is to define the microscopic space\&time-integrated current $\vq$ in the bulk of the lattice during a given observation time interval $[0,t]$. In few words, every time a particle jumps between two bulk neighboring sites along the $\alpha$-direction, $\alpha=x,y$, we add or subtract one to the corresponding $\alpha$-component of the integrated current. In this way, the space\&time-averaged current vector is defined as
\be
\vq=\frac{1}{t}\Big(\frac{1}{L-1}(Q^{+,x}_t-Q^{-,x}_t),\frac{1}{L}(Q^{+,y}_t-Q^{-,y}_t)\Big)
\ee
where $Q^{\pm,\alpha}_t$ are the total number of particle jumps in the $\pm \alpha$-direction, $\alpha=x,y$, in a given microscopic time interval $[0,t]$. Recall that as we are considering the contributions of all the bulk pairs we have to divide the current by $(L-1)$ if the jump occurs in the $x$-direction or by $L$ if it occurs in the $y$-direction in order to count the number of particles that traverses the system per unit \emph{area} and unit time. The empirical averaged current obeys a large deviation principle with large deviation function
\be
G_L(\vq)=\lim_{t\to \infty}\frac{1}{t}\log P(\vq) \, ,
\ee
and its scaled cumulant generating function (SCGF) is defined as
\be
\label{scgf}
\mu_L(\vlamb)=\lim_{t\to \infty}\frac{1}{t}\log \la e^{t \vlamb \cdot \vq}\ra.
\ee
It is then easy to show \cite{RakosHarris} that the SCGF is linked to the spectral properties of a modified or tilted Hamiltonian ${\hat H}$. In particular, $\la e^{t \vlamb \cdot \vq}\ra = \la e^{-{\hat H} t }\ra$, where the new operator ${\hat H}$ is obtained by multiplying the terms of $H$ corresponding to bulk particle transitions by $e^{\pm \lam_x/(L-1)}$ for jumps in the $\pm x$-direction and by $e^{\pm \lam_y/L}$ for jumps in the $\pm y$-direction. This modified Hamiltonian hence reads 
\bea
\label{modH}
-\hat H&=&\sum_{j=1}^L\Big\{\alpha (a_{j,1}^+-1)+\gamma (a_{j,1}^--d_{j,1})+\delta (a_{j,L}^+-1)+\beta (a_{j,L}^--d_{j,L})\nonumber\\
&+&\sum_{i=1}^{L-1} h_x (e^{\frac{\lam_x}{L-1}}a_{j,i}^-a_{j,i+1}^+ -d_{j,i}) + h_x (e^{\frac{-\lam_x}{L-1}}a_{j,i}^+a_{j,i+1}^- -d_{j,i+1})\nonumber\\
&+&\sum_{i=1}^L h_y (e^{\frac{\lam_y}{L}}a_{j,i}^-a_{j+1,i}^+ -d_{j,i}) + h_y (e^{\frac{-\lam_y}{L}}a_{j,i}^+a_{j+1,i}^- -d_{j+1,i})\Big\}.
\eea
Now, assuming that the spectrum of ${\hat H}$ is gapped and introducing the associated spectral decomposition, we can write
\be
\label{decomp}
\la e^{t \vla \cdot \vq}\ra=\la 1|e^{-{\hat H} t }| P_0\ra=\sum_k \la 1| \phi_k\ra \la \phi_k |P_0\ra e^{-\epsilon_k(\vla) t} \xrightarrow[]{t\gg 1} \la 1| \psi \ra \la \psi |P_0\ra e^{-\epsilon_0(\vla) t}
\ee
where $|\psi\ra$ and $\la \psi |$ are the right and left eigenvectors of ${\hat H}$ associated with the lowest eigenvalue $\epsilon_0(\vlamb)$, and $| P_0\ra$ is an arbitrary specific initial particle distribution obeying the normalization condition $\la 1|P_0\ra=1$. If all prefactors in \eqref{decomp} are finite (i.e if $\la 1 |\psi \ra$, $\la \psi |P_0\ra$ and  $\la \psi |\psi \ra$ are finite) one finds, by using \eqref{scgf} and \eqref{decomp}, that 
\be
\label{scgfei}
\mu_L(\vlamb)=-\epsilon_0(\vlamb).
\ee
To compute $\epsilon_0(\vlamb)$, we assume that the unnormalized right eigenvector has a product form similar to  \eqref{statmea}, i.e,
\be
\label{pfright}
|\psi\ra=|\psi_{1,1}) \otimes |\psi_{1,2}) \otimes \cdots \otimes |\psi_{L,L})
\ee
where $|\psi_{j,i})$ is the vector for the $(j,i)$-site, i.e, $|\psi_{j,i})= \sum_{n_{ji}}\psi^{\text{right}}_{j,i}(n_{ji})|n_{ji})$, whose components are $\psi^{\text{right}}_{j,i}(n_{ji})={\hat z}_{j,i}^{n_{ji}}\prod_{k=1}^{n_{ji}} f(k)^{-1}$ with ${\hat z}_{j,i}$ some modified fugacities still unknown. With the product form \eqref{pfright} one can readily check that 
\be
\label{crearig}
a_{j,i}^+|\psi\ra = {\hat z}_{j,i}^{-1} d_{j,i} |\psi\ra,
\ee
\be
\label{anrig}
a_{j,i}^-|\psi\ra = {\hat z}_{j,i} |\psi\ra.
\ee
Using these equations we get that
\bea
\label{rightvec}
-{\hat H}| \psi \ra &=& \sum_{j=1}^{L} \Big\{-(\alpha+\delta-(\gamma {\hat z}_{j,1}+\beta {\hat z}_{j,L})) \nonumber \\
&+&\sum_{i=2}^{L-1} z_{j,i}^{-1} d_{j,i} \Big[ 
{\hat z}_{j,i+1} h_x e^{\frac{-\lam_x}{L-1}}-{\hat z}_{j,i} (2h_x+h_y(1-e^{\frac{\lam_y}{L}}) + h_y (1-e^{\frac{-\lam_y}{L}}) )  +{\hat z}_{j,i-1} h_x e^{\frac{\lam_x}{L-1}} \Big] \nonumber \\
&+& {\hat z}_{j,1}^{-1}d_{j,1} \Big(  
{\hat z}_{j,2} h_x e^{\frac{-\lam_x}{L-1}}-{\hat z}_{j,1} (h_x+\gamma+h_y(1-e^{\frac{\lam_y}{L}}) + h_y (1-e^{\frac{-\lam_y}{L}}) )  +\alpha 
\Big) \nonumber \\
&+& {\hat z}_{j,L}^{-1} d_{j,L} \Big(
{\hat z}_{j,L-1} h_x e^{\frac{\lam_x}{L-1}}-{\hat z}_{j,L} (\beta+h_x+h_y(1-e^{\frac{\lam_y}{L}}) + h_y (1-e^{\frac{-\lam_y}{L}}) )  +\delta
\Big)\Big\}|\psi\ra. \nonumber \\
\eea
It is clear that if $|\psi \ra$ is a right eigenvector, the coefficients that multiply the matrix $d$ must vanish. In this way, we can compute the components of the right eigenvector fugacities by solving the following recurrence relation
\be
\label{bulkrecrig}
{\hat z}_{i+1} h_x e^{\frac{-\lam_x}{L-1}}-{\hat z}_i (2h_x+h_y(1-e^{\frac{\lam_y}{L}}) + h_y (1-e^{\frac{-\lam_y}{L}}) )  +{\hat z}_{i-1} h_x e^{\frac{\lam_x}{L-1}}=0
\ee
with boundary conditions
\be
\label{bc1rig}
{\hat z}_{2} h_x e^{\frac{-\lam_x}{L-1}}-{\hat z}_{1} (h_x+\gamma+h_y(1-e^{\frac{\lam_y}{L}}) + h_y (1-e^{\frac{-\lam_y}{L}}) )  +\alpha=0
\ee
\be
\label{bc2rig}
{\hat z}_{L-1} h_x e^{\frac{\lam_x}{L-1}}-{\hat z}_L (\beta+h_x+h_y(1-e^{\frac{\lam_y}{L}}) + h_y (1-e^{\frac{-\lam_y}{L}}) )  +\delta=0.
\ee
Notice that in the previous equations we have made use of the periodic boundary conditions to argue that fugacities are invariant in the $y$-direction, ${\hat z}_{j,i}={\hat z}_i$, $\forall j$. Equations \eqref{bulkrecrig}-\eqref{bc2rig} can be solved exactly with a computer to get the fugacity of the right eigenvector for each column ${\hat z}_i$  ($i\in[1,L]$), but the expressions obtained are too cumbersome to write them explicitly here. Thus, from eqs. \eqref{rightvec}-\eqref{bc2rig}, we get the lowest eigenvalue of ${\hat H}$
\be
{\hat H}| \psi \ra= L \Big( \alpha+\delta-(\gamma {\hat z}_1+\beta {\hat z}_L) \Big)| \psi \ra=\epsilon_0(\vla) | \psi \ra \, ,
\ee
from which that SCGF in \eqref{scgfei} follows as
\be
\label{murig}
\mu_L(\vla)=-L\Big( \alpha+\delta-(\gamma {\hat z}_1+\beta {\hat z}_L) \Big) \, ,
\ee
with ${\hat z}_{1,L}$ explicitly given in terms of the solution of recurrence (\ref{bulkrecrig})-(\ref{bc2rig}).

\subsection{Microscopic optimal density profiles for the $2d$ ZRP}
In order to compute the mean density in each site, we need both the right and left dominant eigenvectors associated to a given current fluctuation. For the left eigenvector, we assume again a product form similar to eq. \eqref{pfright}, i.e \cite{RakosHarris}
\be
\label{pfleft}
\la \psi|=( \psi_{1,1}| \otimes (\psi_{1,2}| \otimes \cdots \otimes (\psi_{L,L}|
\ee
where $(\psi_{j,i}|$ is the vector for the $(j,i)$-site, i.e, $(\psi_{j,i}|= \sum_{n_{ji}} \psi^{\text{left}}_{j,i}(n_{ji})  (n_{ji}|$, whose components are $\psi^{\text{left}}_{j,i}(n_{ji})={\tilde z}_{j,i}^{n_{ji}}$ with ${\tilde z}_{j,i}$ some modified fugacities to be determined below. With the product form \eqref{pfleft} one can readily check that 
\be
\label{crealef}
\la \psi | a_{j,i}^+=\la \psi | {\tilde z_{j,i}},
\ee
\be
\label{anlef}
\la \psi | a_{j,i}^-=\la \psi | {\tilde z}_{j,i}^{-1}d_{j,i}.
\ee
Using these equations we get that 
\bea
-\la \psi |{\hat H} &=& \la \psi |  \sum_{j=1}^{L}\Big\{ -(\alpha+\delta-(\alpha {\tilde z}_{j,1}+\delta {\tilde z}_{j,L})) \nonumber \\
&+& \sum_{i=2}^{L-1} z_{j,i}^{-1} d_{j,i} \Big[ 
{\tilde z}_{j,i+1} h_x e^{\frac{\lam_x}{L-1}}-{\tilde z}_{j,i} (h_x+h_x+h_y(1-e^{\frac{\lam_y}{L}}) + h_y (1-e^{\frac{-\lam_y}{L}}) )  +{\tilde z}_{j,i-1} h_x e^{\frac{-\lam_x}{L-1}} \Big] \nonumber \\
&+& {\tilde z}_{j,1}^{-1}d_{j,1} \Big(  
{\tilde z}_{j,2} h_x e^{\frac{\lam_x}{L-1}}-{\tilde z}_{j,1} (h_x+\gamma+h_y(1-e^{\frac{\lam_y}{L}}) + h_y (1-e^{\frac{-\lam_y}{L}}) )  +\gamma 
\Big) \nonumber \\
&+& {\tilde z}_{j,L}^{-1} d_{j,L} \Big(
{\tilde z}_{j,L-1} h_x e^{\frac{-\lam_x}{L-1}}-{\tilde z}_{j,L} (\beta+h_x+h_y(1-e^{\frac{\lam_y}{L}}) + h_y (1-e^{\frac{-\lam_y}{L}}) )  +\beta
\Big)\Big\}. \nonumber \\
\eea
As before, the coefficients multiplying matrix $d$ must vanish, resulting in the following recurrence relation for the left eigenvector fugacities (where we have considered again that ${\tilde z}_{j,i}={\tilde z}_i$, $\forall j$)
\be
\label{bulkreclef}
{\tilde z}_{i+1} h_x e^{\frac{\lam_x}{L-1}}-{\tilde z}_i (2h_x+h_y(1-e^{\frac{\lam_y}{L}}) + h_y (1-e^{\frac{-\lam_y}{L}}) )  +{\tilde z}_{i-1} h_x e^{\frac{-\lam_x}{L-1}}=0
\ee
with boundary conditions
\be
\label{bc1lef}
{\tilde z}_{2} h_x e^{\frac{\lam_x}{L-1}}-{\tilde z}_1 (h_x+\gamma+h_y(1-e^{\frac{\lam_y}{L}}) + h_y (1-e^{\frac{-\lam_y}{L}}) )  +\gamma =0
\ee
\be
\label{bc2lef}
{\tilde z}_{L-1} h_x e^{\frac{-\lam_x}{L-1}}-{\tilde z}_L (\beta+h_x+h_y(1-e^{\frac{\lam_y}{L}}) + h_y (1-e^{\frac{-\lam_y}{L}}) )  +\beta=0.
\ee
By solving eqs. \eqref{bulkreclef}-\eqref{bc2lef} we get the fugacity of the left eigenvector for each column, ${\tilde z}_i$  ($i\in[1,L]$). Moreover, the SCGF can be equivalently written in terms of these fugacities as
\be
\label{mulef}
\mu_L(\vla)=-L\Big( \alpha+\delta-(\alpha {\tilde z}_1+\delta {\tilde z}_L) \Big).
\ee
One can check that as $\vla \to 0$, $\la \psi |\to \la 1|$, $|\psi \ra\to |P^*\ra$ and $\mu_L(\vla)\to 0$, as expected. Finally, we can compute the microscopic optimal profiles associated to a current fluctuation (parametrized via $\vla$) by averaging the mean occupation number in each column $i$ over the right and left dominant eigenvectors, once normalized, i.e. 
\be
\rho_{j,i}\equiv \la n_{ji} \ra= \frac{\la \psi|{\hat n}_{ji}|\psi \ra}{\la \psi| \psi \ra}=\frac{({\tilde z}_i {\hat z}_i)^{n_{ji}} \prod_{k=1}^{n_{ji}}f(k)^{-1}} {\sum_{n_{ji}=0}^{\infty} ({\tilde z}_i {\hat z}_i)^{n_{ji}} \prod_{k=1}^{n_{ji}}f(k)^{-1}}={\bar z}_i\frac{\partial \log {\bar Z}_i}{\partial {\bar z}_i }
\ee
where ${\bar z}_i\equiv {\tilde z}_i {\hat z}_i$
and 
\be
{\bar Z}_i\equiv \sum_{n=0}^{\infty} {\bar z}_i^n \prod_{k=1}^{n} f(k)^{-1}.
\ee
As expected, the mean occupation number on site $(j,i)$ just depends on ${\bar z}_i$, so the microscopic density profile associated to a given current fluctuation exhibits structure only along the gradient direction, in agreement with general MFT predictions in the main text.

On the other hand, in this work we consider two different interaction functions $f(k)$. The first one is a constant $f(k)=1$, and corresponds to a $2d$ ZRP with effective attractive interaction between particles at each lattice site. The associated optimal density profile is
\be
\label{prof-zrp}
\rho_i=\frac{{\bar z}_i}{1-{\bar z}_i}.
\ee
In this case, the reservoir fugacities in terms of the injection and extraction rate are given by $z_{1}=\alpha/\gamma$ and $z_{L}=\delta/\beta$ and the reservoirs densities by $\rho_L=\frac{\alpha/\gamma}{1-\alpha/\gamma}$ and $\rho_R=\frac{\delta/\beta}{1-\delta/\beta}$. The parameters chosen in the main text for the isotropic ZRP case (with $h_x=1/2$, $h_y=1/2$), whose results are displayed in Fig.~\ref{fig1} and Fig.~\ref{fig4}, are $\alpha=1/4$, $\gamma=1/2$, $\delta=1/22$ and $\beta=1/2$. These parameters correspond to $\rho_L=1$ and $\rho_R=0.1$. The same parameters are chosen for the anisotropic ZRP case studied in Appendix B (with $h_x=1/2$, $h_y=1$), see Fig.~\ref{fig3} there. The second interaction function that we consider is $f(k)=k$ and corresponds to a $2d$ fluid of independent random walkers (RW), giving rise to the following optimal microscopic density profile
\be
\label{prof-rw}
\rho_i={\bar z}_i.
\ee
The reservoirs fugacities are given by $z_{1}=\alpha/\gamma$ and $z_{L}=\delta/\beta$, and the reservoirs densities are now $\rho_L=\alpha/\gamma$ and $\rho_R=\delta/\beta$. The parameters chosen in the isotropic RW case studied in Appendix B (with $h_x=1/2$, $h_y=1/2$), see Fig.~\ref{fig5}, are $\alpha=1$, $\gamma=1/2$, $\delta=1$, $\beta=1$, which correspond to $\rho_L=2$ and $\rho_R=1$.

\subsection{Comparing microscopic and macroscopic results}
In order to compare the previous microscopic results above with macroscopic fluctuation theory predictions we need to perform a diffusive scaling on the microscopic results. This consists in the following transformations of space and time: $x=i/L$, $y=j/L$ and $\tau=t/L^2$, where $i$, $j$, $t$ are the microscopic space and time variables and $x$, $y$ and $\tau$ the macroscopic ones. Applying this diffusive scaling, the macroscopic SCGF reads
\be
\mu(\vla)=\lim_{L\to \infty}\frac{\mu_L(\vla)}{L^{d-2}} \, .
\ee
Therefore, in $d=2$ we have $\mu(\vla)=\lim_{L\to \infty} \mu_L(\vla)$, with $\mu_L(\vla)$ given by eq. \eqref{murig}. Then for every $\vla^*=(\lam_x^*,\lam_y^*)$ we can calculate the current large deviation function knowing that 
\be
G(\vJJ)=\max_{\vla}[\mu(\vla)-\vla \cdot \vJJ]=\mu(\vla^*)-\vla^* \cdot \vJJ
\ee
where $\vJJ=(J_x,J_y)=\left(\left. \frac{\partial \mu(\vla)}{\partial \lam_x}\right|_{\vla=\vla^*},\left. \frac{\partial \mu(\vla)}{\partial \lam_y}\right|_{\vla=\vla^*} \right)$. Finally, the optimal macroscopic profile $\rho(x)$ (with $x\in [0,1]$) is nothing but the microscopic optimal profile $\rho_i$ with $x=\frac{i}{L}$ and $\rho_i$ given by eqs. \eqref{prof-zrp} or \eqref{prof-rw} depending on the interaction function at play. As a crosscheck of our results, note that the macroscopic and microscopic optimal profiles obtained for the ZRP with $f(k)=1$ for the angles $\phi=0,\pi$ are the same of those obtained in Ref. \cite{Schutz3} for the one-dimensional symmetric case.

\newpage

\twocolumngrid

\section{Some additional results}
\label{apB}

\begin{figure}[b!]
\includegraphics[width=8.5cm]{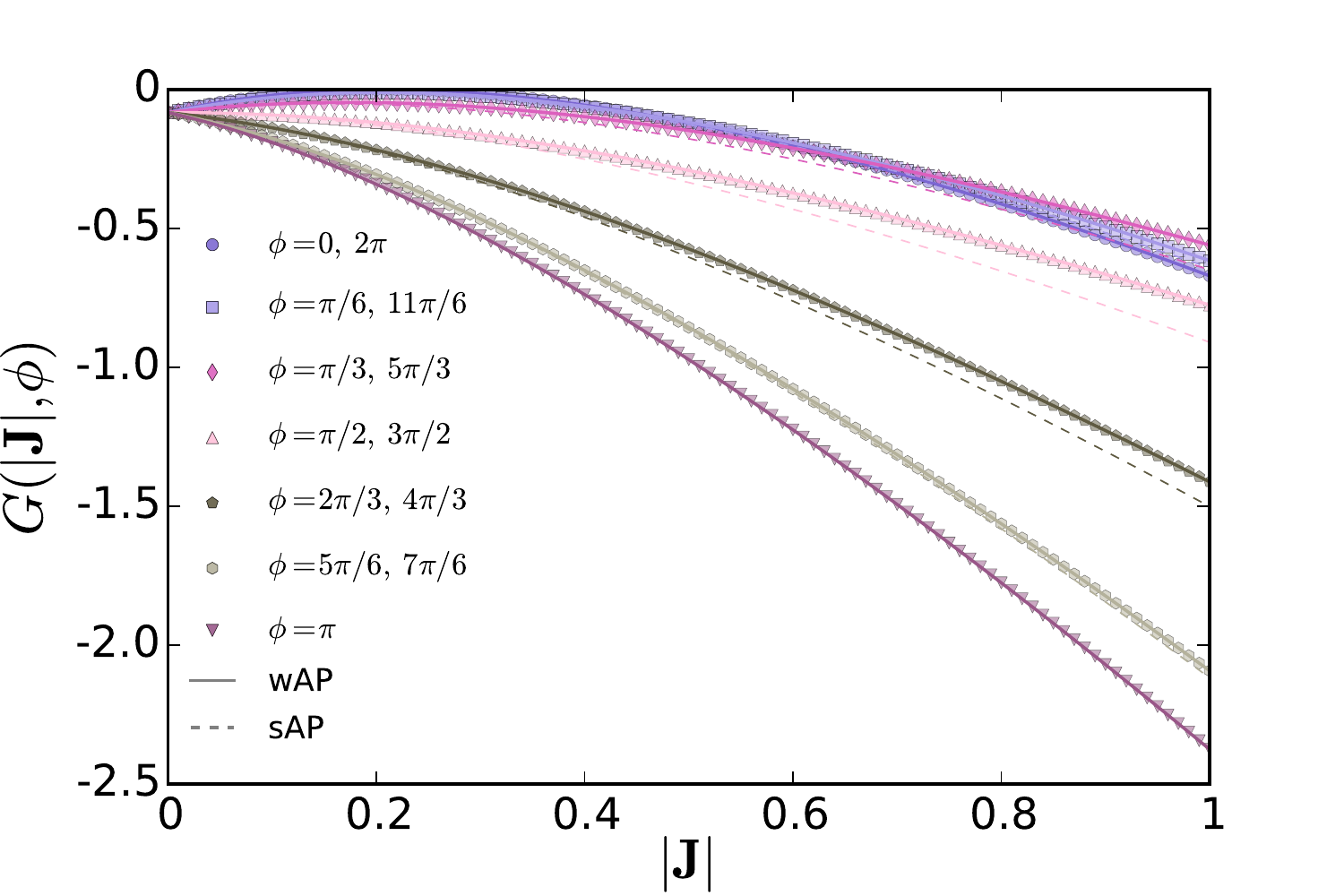}
\includegraphics[width=8.5cm]{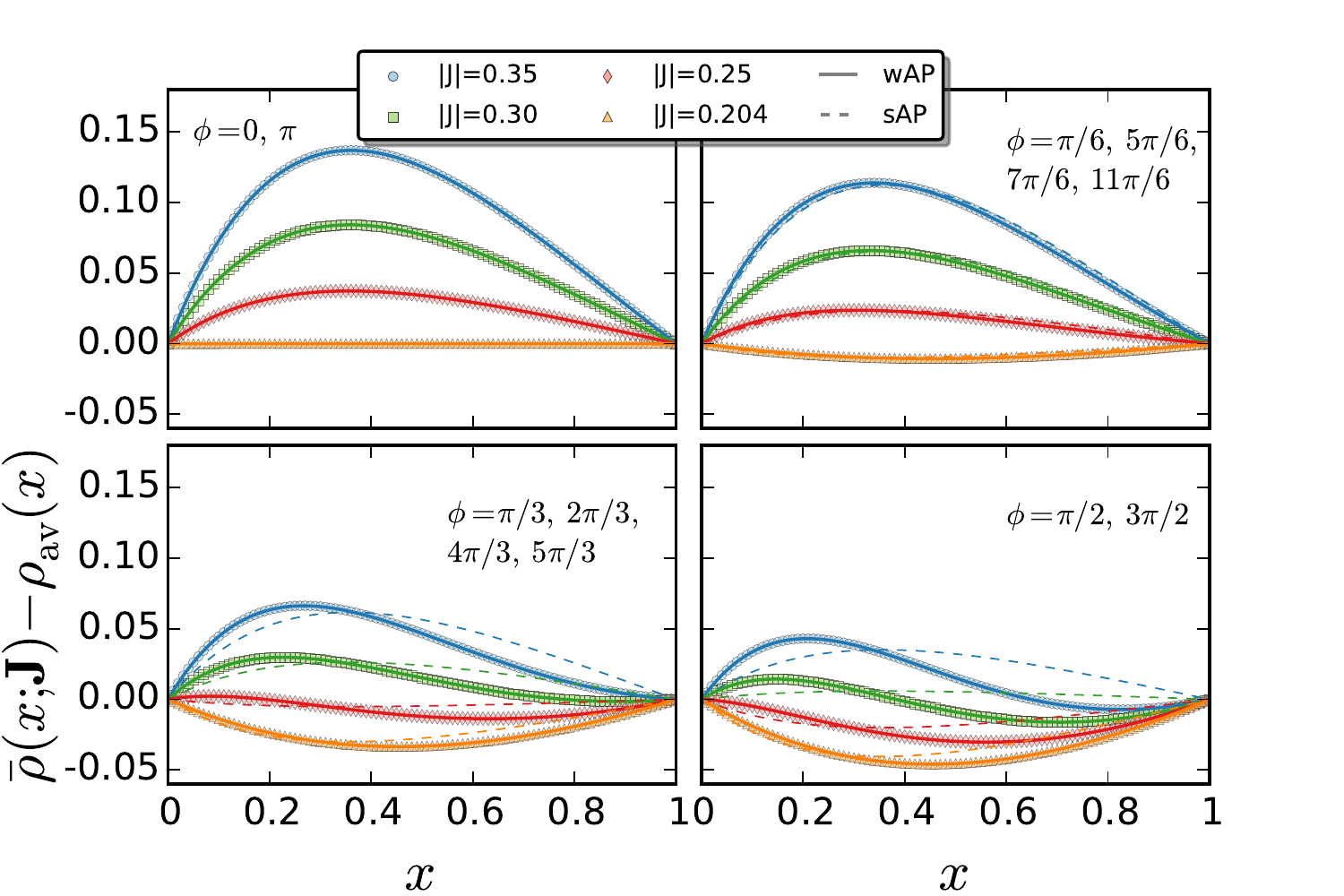}
\caption{\small (Color online) Top: Current LDF for the \emph{anisotropic} ZRP ($h_x=1/2$, $h_y=1$) with $\rho_L=1$ and $\rho_R=0.1$, as a function of $|\vJJ|$ for different angles $\phi=\tan^{-1}(J_y/J_x)$. Bottom: Excess optimal density profiles for different $|\vJJ|$ and $\phi$. Symbols stand for exact matrix computations for $L=10^5$, while solid (dashed) lines represent wAP (sAP) predictions. 
\label{fig3}
}
\end{figure}

In this Appendix we provide additional data which support our conclusions in the main text. In particular, we report further exact microscopic results obtained by applying the quantum Hamiltonian formalism of Appendix A to the Zero Range Process (ZRP) described in the main text \cite{zrp}, both in the isotropic and anisotropic cases, and to a fluid of random walkers (RW model).

\begin{figure}[b!]
\includegraphics[width=8.5cm]{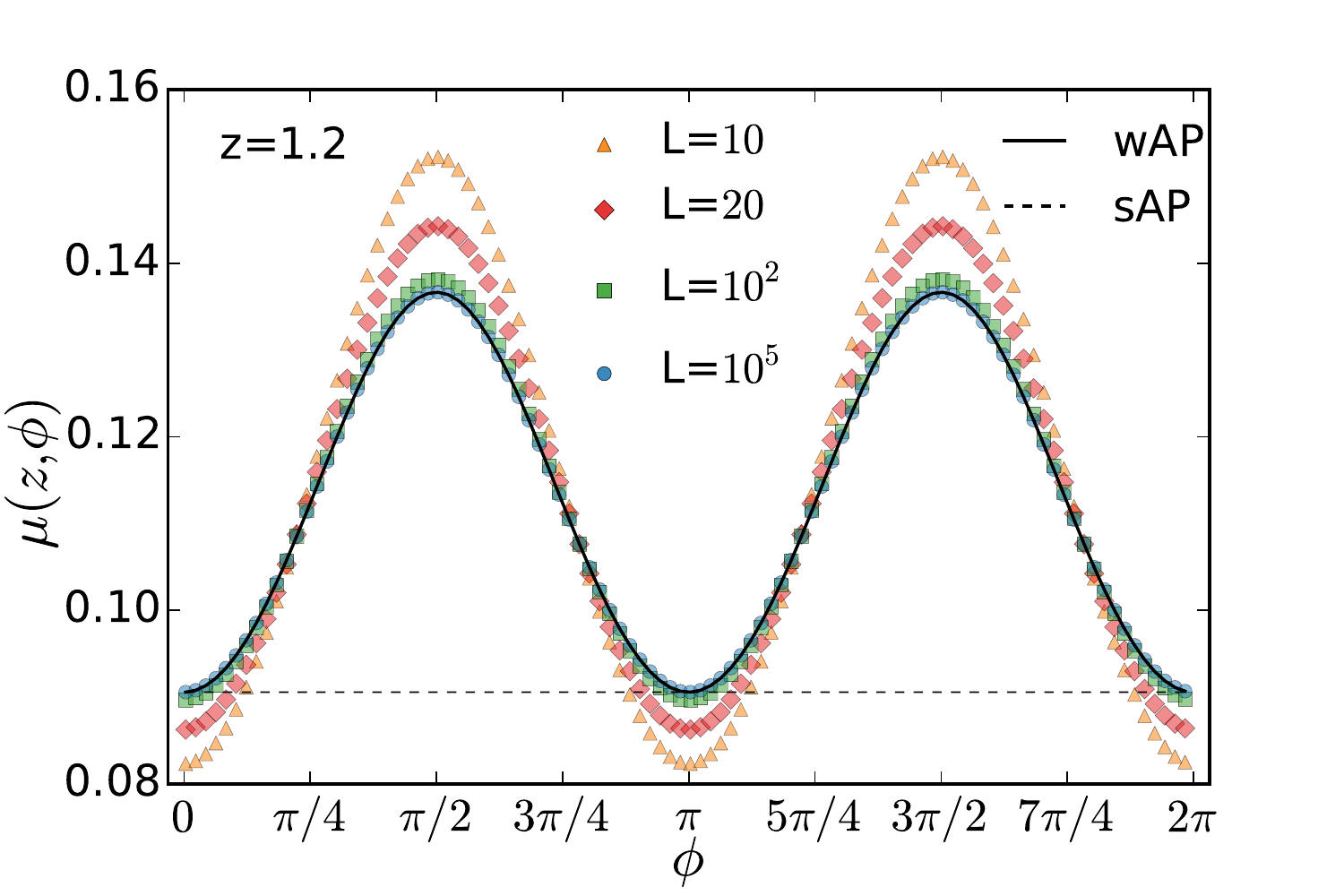}
\caption{\small (Color online) Legendre transform of the current LDF of the isotropic ZRP $(h_x=1/2,\, h_y=1/2)$ as a function of $\phi$ for $z\equiv |\zz|=1.2$, $\rho_L=1$, $\rho_R=0.1$ and different system sizes $L$. Symbols stand for exact matrix computations, while solid (dashed) lines represent wAP (sAP) predictions. Convergence to the wAP prediction as $L$ increases is apparent, similarly to the behavior observed for the KMP result in Fig.~2 of the main text.
\label{fig4}
}
\end{figure}

We first focus on the effect of anisotropy on the current LDF and the associated optimal density profiles. Fig.~\ref{fig3} shows $G(\vJJ)$ (top) and the optimal density profiles $\bar\rho(x;\vJJ)$ (bottom, after subtracting the steady-state profile $\rho_\text{av}(x)$) for the anisotropic ZRP with jump rates $(h_x=1/2,\, h_y=1)$, $L=10^5$, and boundary densities $\rho_L=1$ and $\rho_R=0.1$. Similarly to the results in the main text, wAP predictions perfectly fit the exact microscopic results derived within the matrix approach of Appendix A. On the other hand, theoretical curves based on the sAP fail to correctly predict the shape of $G(\vJJ)$ and the associated optimal density profiles, except for currents $\vJJ=(\Jpar,0)$ alligned with the gradient direction, where both wAP and sAP predictions converge as proven in the main text. In any case, it is interesting to note that optimal density profiles responsible of a given current fluctuation $\vJJ$ are typically different from the average, steady-state density profile, see bottom panel in Fig.~\ref{fig3}. This general observation, common to all studied models, stems from the (typically nonlinear) dependence of the diffusivity and mobility transport coefficients on the local density field. 

\begin{figure}[b!]
\includegraphics[width=8.5cm]{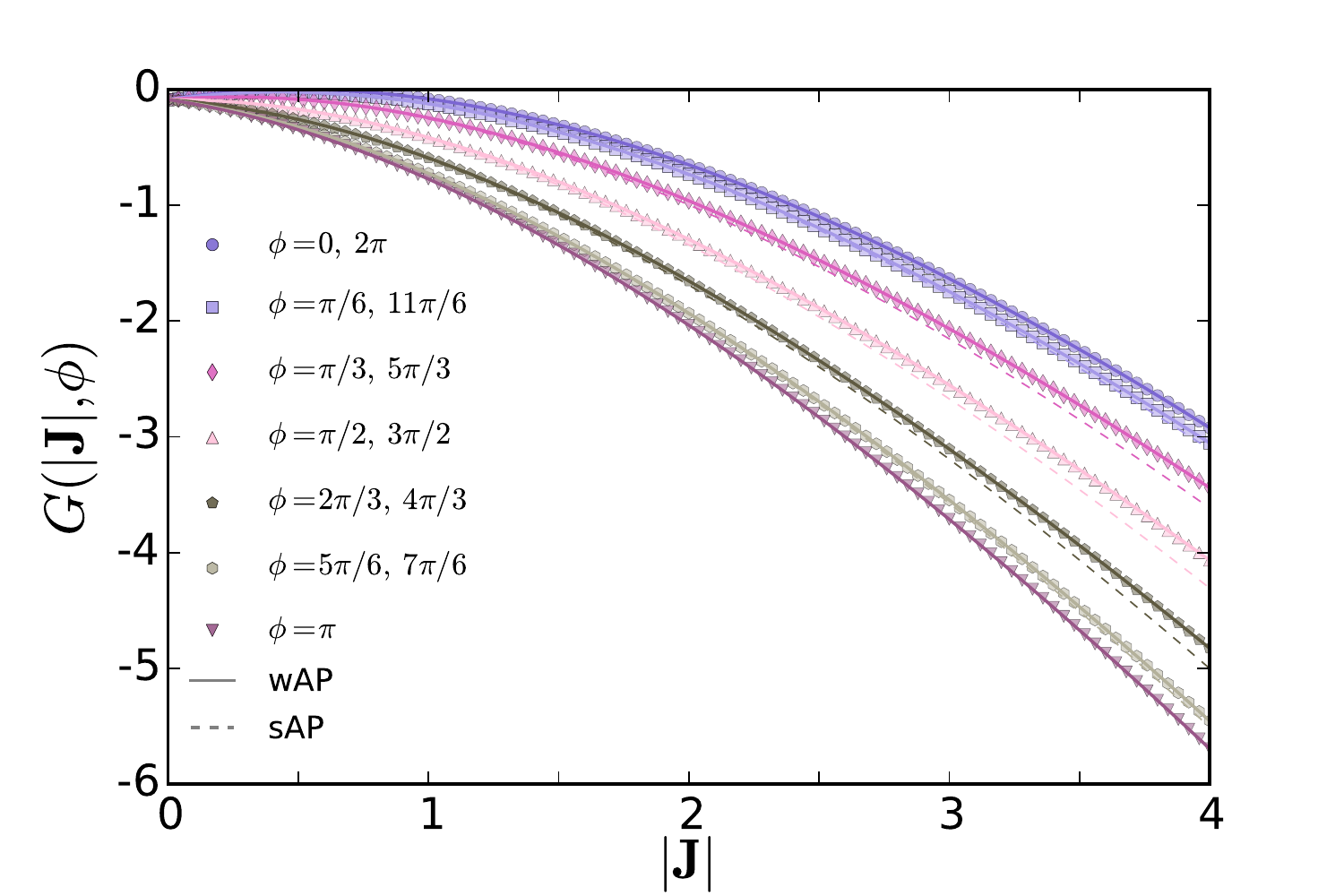}
\includegraphics[width=8.5cm]{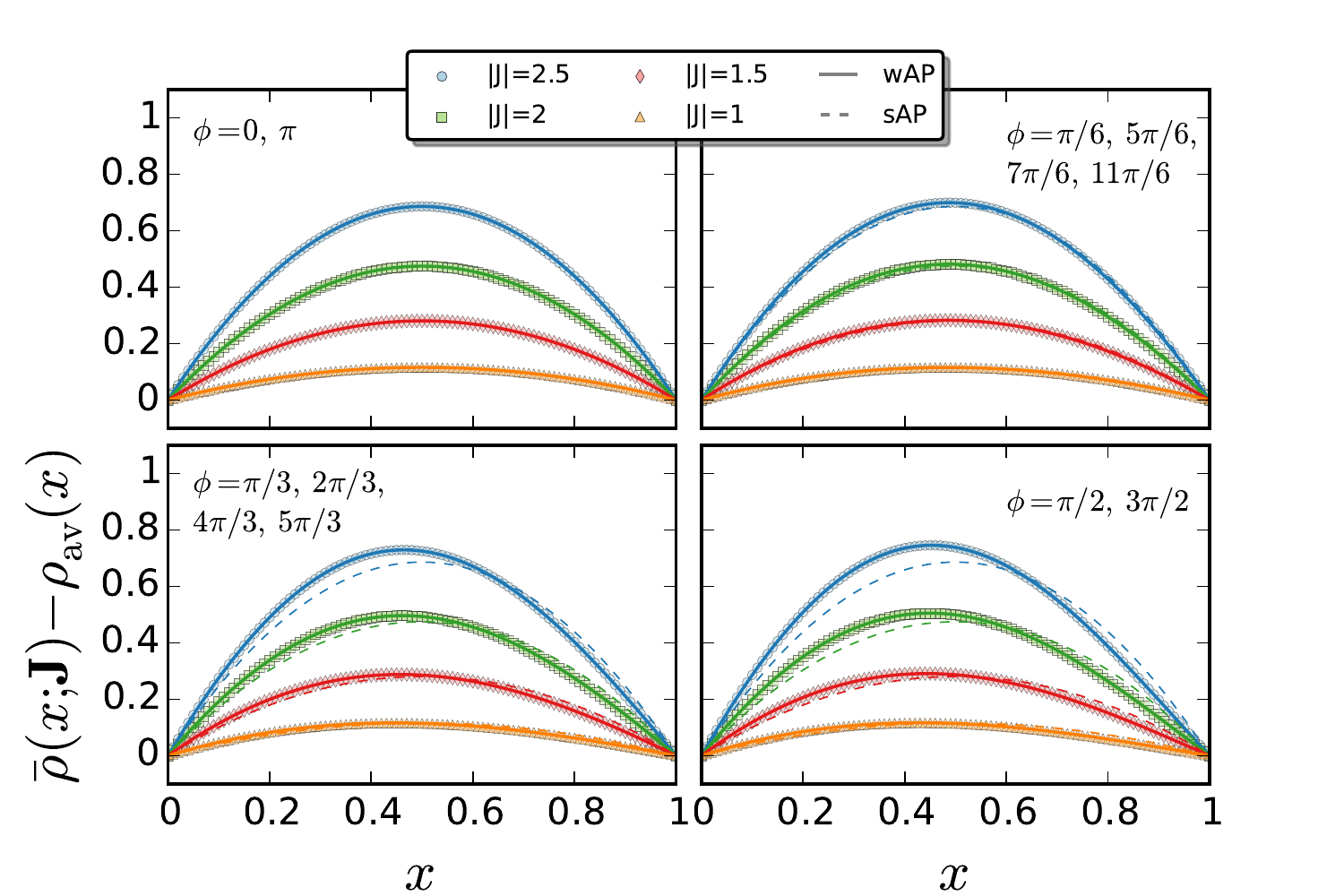}
\caption{\small (Color online) Top: Current LDF for the isotropic RW model $(h_x=1/2,\, h_y=1/2)$ as a function of $|\vJJ|$ for different angles $\phi=\tan^{-1}(J_y/J_x)$ and , $\rho_L=2$, $\rho_R=1$. Bottom: Excess optimal density profiles for different $|\vJJ|$ and $\phi$. Symbols stand for exact matrix computations for $L=10^5$, while solid (dashed) lines represent wAP (sAP) predictions. 
\label{fig5}
}
\end{figure}

The power of the quantum Hamiltonian formalism for the master equation, when combined with the factorization property of the ZRP \cite{zrp}, allows us to study its current statistics not only for very large lattice linear sizes $L=10^5$, but also to understand the role of finite-size corrections from a microscopic point of view. We exploit now this possibility in order to compare the finite-size behavior of the ZRP with the more complex KMP model studied in the main text \cite{kmp}, for which reliable data for current statistics can be obtained only for relatively small system sizes via rare-event Monte Carlo simulation techniques \cite{weJSP,sim1,sim2,sim3,sim4}. Fig.~\ref{fig4} shows for the isotropic ZRP $(h_x=1/2,\, h_y=1/2)$ the Legendre-Fenchel transform of the current LDF, 
\be
\mu(\zh)=\max_\vJJ[G(\vJJ) + \zh\cdot \vJJ] \, , \nonumber
\ee
for a fixed value of $z=|\zz|$, with $\zz\equiv \zh+\vecep$ and $\vecep=\frac{1}{2}\ln[\rho_L(1+\rho_R)/(\rho_R(1+\rho_L))]$, as a function of the current angle $\phi=\tan^{-1}(J_y/J_x)$ for $\rho_L=1$, $\rho_R=0.1$, and increasing values of $L$, together with the wAP and sAP predictions. As described in the main text, while sAP predicts no angular dependence for $\mu(\zh)$, the wAP does predicts a double-bump structure, which is fully confirmed in exact microscopic calculations, see Fig.~\ref{fig4}. Moreover, data points for small $L$ converge towards the wAP curve as $L$ increases, very much like the results obtained for the KMP model of heat conduction, see Fig.~2 in the paper. This observation supports our analysis and conclusions for the KMP model, which strongly suggest that the weak additivity principle is indeed correct for sufficiently large system sizes.

\begin{figure}
\includegraphics[width=8.5cm]{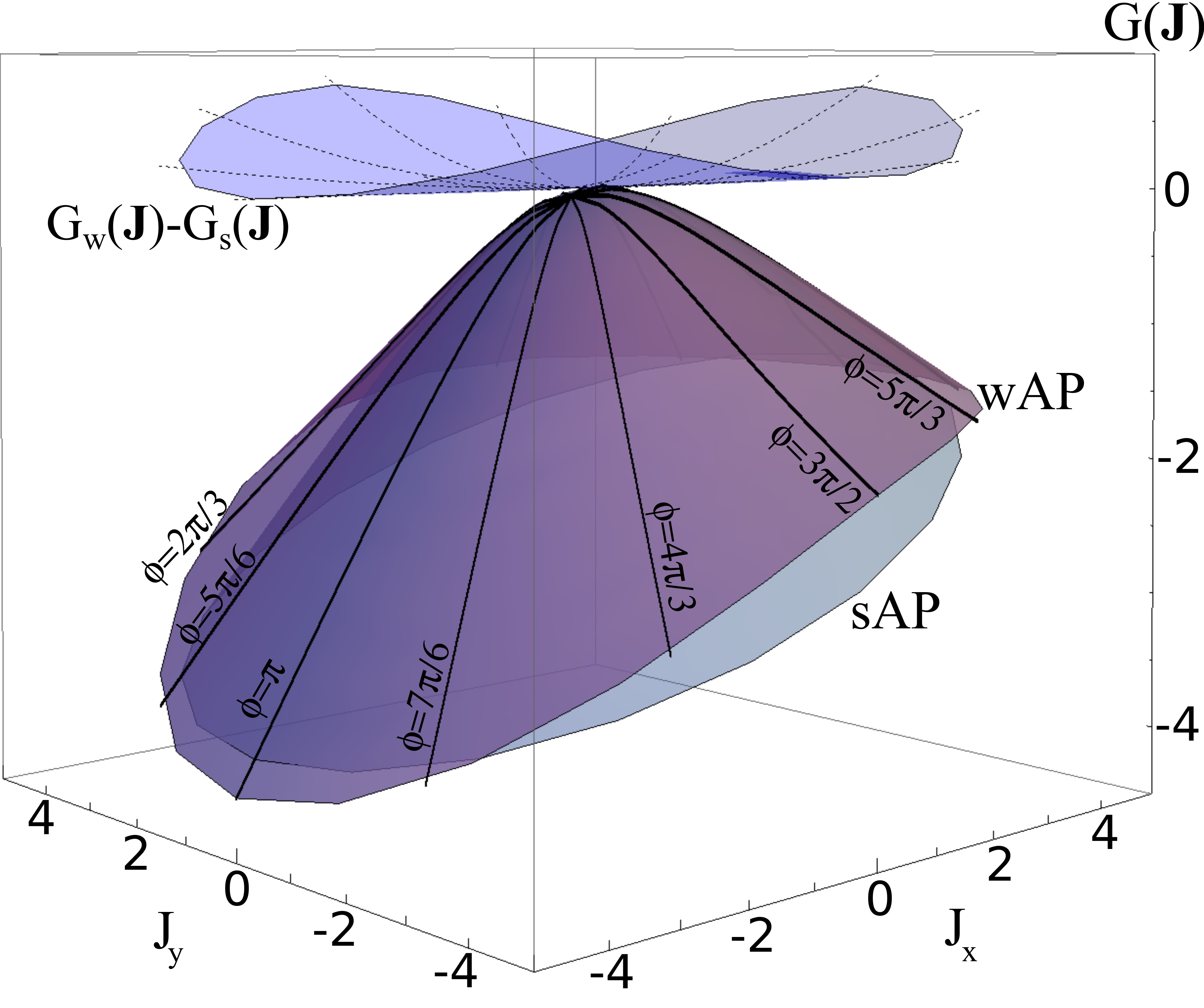}
\caption{\small (Color online) MFT predictions for the current LDF of the KMP model under both the wAP and sAP conjectures. Boundary densities for this plot are $\rho_L=2$ and $\rho_R=1$. Clearly, $G_\wap(\vJJ)$ lies above $G_\sap(\vJJ)$ $\forall \vJJ$, except for current fluctuations along the gradient direction, $\vJJ=(\Jpar,0)$ $\forall \Jpar$, where both solutions yield the same result, as demonstrated in the main text.
\label{fig6}
}
\end{figure}

To end this section, we apply the quantum Hamiltonian formalism to another stochastic lattice model, a fluid of random walkers \cite{zrp}. The RW model can be seen as a variant of the ZRP with an interaction function $f(n)=n$. Such a linear interaction function implies that the probability for \emph{each particle} to jump to a nearby site is independent of the population of the departure site, so particles behave as independent random walkers in $d=2$. At the macroscopic level, the RW model is characterized by transport coefficients with components $D_\alpha(\rho)=h_\alpha$ and $\sigma_\alpha(\rho)=2h_\alpha\rho$, so when coupled to boundary reservoirs at densities $\rho_{L,R}$ along the $x$-direction, with $\rho_L\ne \rho_R$, the RW fluid develops a linear stationary density profile $\rho_\text{av}(x)=\rho_L+x(\rho_R-\rho_L)$ similar to that of the KMP model. Note however that the fluctuating behavior of both models is quite different because of their different mobilities. Fig.~\ref{fig5} shows our results for $G(\vJJ)$ (top) and $\bar\rho(x;\vJJ)$ (bottom) in the RW model, as obtained from the matrix method for $L=10^5$ and compared with wAP and sAP theoretical results. Interestingly differences between wAP and sAP curves are not as pronounced as before (due to the relatively \emph{weak} dependence of the transport coefficients on $\rho$ for the RW model), but still the wAP offers correct predictions while the sAP fails for current fluctuations with components orthogonal to the gradient direction. All together, these results and those reported in the main text clearly demonstrate that the weak additivity principle yields the correct predictions for the current statistics of a broad class of $d$-dimensional interacting particle systems. 

Finally, Fig.~\ref{fig6} displays the MFT prediction for the full current LDF of the 2d KMP model for $\rho_L=2$ and $\rho_R=1$ under both the wAP and sAP conjectures. Clearly the wAP current LDF improves over the sAP prediction for all current fluctuations, as proven in general in the main text on the basis of reverse H\"older's inequality.


\begin{thebibliography}{99}
\bibitem{note1} The physical picture behind this hypothesis corresponds to a system that, after a short transient time at the beginning of the large deviation event (microscopic in the diffusive timescale $\tau$), settles into a time-independent state with structured density and current fields (which can be different from the stationary ones) such that the empirical, space\&time-averaged current equals $\vJJ$. This behavior is expected to minimize the cost of a fluctuation at least for moderate deviations from the average behavior.
\bibitem{rhoav} The ZRP steady-state density profile is $\rho_\text{av}(x)=[\rho_L(1+\rho_R)-x(\rho_L-\rho_R)]/[1+\rho_R+x(\rho_L-\rho_R)]$ in this case.
\end{thebibliography}
\end{document}